# Leveraging SeNet and ResNet Synergy within an Encoder-Decoder Architecture for Glioma Detection

Pandiyaraju V*[1], Shravan Venkatraman[2], Abeshek A[3], Pavan Kumar S[4], Aravintakshan S A[5]

[1,2,3,4,5,] School of Computer Science and Engineering, Vellore Institute of Technology, Chennai, Tamil Nadu 600127, India

*Abstract:*

Brain tumors are abnormalities that can severely impact a patient's health, leading to life-threatening conditions such as cancer. These can result in various debilitating effects, including neurological issues, cognitive impairment, motor and sensory deficits, as well as emotional and behavioral changes. These symptoms significantly affect a patient's quality of life, making early diagnosis and timely treatment essential to prevent further deterioration. However, accurately segmenting the tumor region from medical images, particularly MRI scans, is a challenging and time-consuming task that requires the expertise of radiologists. Manual segmentation can also be prone to human errors. To address these challenges, this research leverages the synergy of SeNet and ResNet architectures within an encoder-decoder framework, designed specifically for glioma detection and segmentation. The proposed model incorporates the power of SeResNet-152 as the backbone, integrated into a robust encoder-decoder structure to enhance feature extraction and improve segmentation accuracy. This novel approach significantly reduces the dependency on manual tasks and improves the precision of tumor identification. Evaluation of the model demonstrates strong performance, achieving 87% in Dice Coefficient, 89.12% in accuracy, 88% in IoU score, and 82% in mean IoU score, showcasing its effectiveness in tackling the complex problem of brain tumor segmentation.

## 1. INTRODUCTION

Segmentation of brain tumors involves identifying and delineating the affected region from the rest of the brain in MRI images. Once isolated, this region can be used for classification. The process requires expert individuals, such as radiologists, and manually segmenting the area of interest and determining the grade of the abnormality is a tedious and time-consuming task. This manual process may lead to incorrect detection of the tumor region. Therefore, automated segmentation methods are preferred since they reduce the time required and minimize human efforts and errors.

Various strategies involving machine learning and deep learning techniques have been employed for the segmentation of brain tumors from MRI images to reduce the time taken compared to manual methods. Ayachi et al. 2009 [1] utilized a support vector machine (SVM) with T1-weighted and T2-weighted images for brain tumor segmentation. Texture features of first and second order, along with intensities, were adopted, and a feature vector was computed using the MIPAV tool. Since this is a supervised machine learning technique, glioma regions were manually validated using active contour models, and this labeled training dataset was used for both training and testing the SVM. Once trained, the SVM classified the input image and extracted the tumor by grouping tumor pixels, resulting in a binary image.

Abdel-Maksoud et al. 2015 [2] proposed KIFCM, a hybrid clustering technique combining advantages of the K-means algorithm—efficient on large datasets—and the Fuzzy C-means algorithm, which retains most information from the actual image, leading to more accurate segmentation. New cluster centers were recomputed by assigning points to the nearest cluster center based on minimum distances. The process continued until convergence criteria were met, ensuring accurate computation of new cluster centers. This was followed by extracting the region of interest using two segmentation techniques: Thresholding and Active contour by level set method. The hybrid clustering method demonstrated improved performance over individual constituent algorithms.

The remainder of the research article is as follows: Section 2 of this article focuses on the related works done to solve similar problems. Section 3 of the research article focuses on the description of the architecture of the proposed model. Section 4 of the article focuses on the implementation and evaluation of the proposed model as well as the results produced and discussion involving performance comparison with models performing similar tasks while Section 5 concludes this research work

## 2. RELATED WORKS

Pereira et al. (2016) [3] introduced two convolutional neural networks (CNNs) for segmenting high-grade (HGG) and low-grade (LGG) gliomas. The HGG architecture is deeper than the LGG counterpart. They employed Xavier initialization to manage activations and gradients, mitigating issues like vanishing gradients. Regularization techniques were used to combat overfitting, and Leaky ReLU was chosen as the activation function. To enforce volumetric constraints, clusters smaller than a predefined threshold, often misclassified as tumors during segmentation, were removed.

Alqazzaz et al. (2019) [4] proposed an automated segmentation algorithm using the SegNet deep learning model on multi-modal MR images. Each SegNet model for FLAIR, T1ce, T1, and T2 modalities consists of an encoder with thirteen convolutional layers, mirroring VGG16's initial layers. These models generate four score maps corresponding to background, edema, enhancing tumor, and necrosis classifications. The highest-value score maps are combined to enhance classification performance. An encoded feature vector is derived from these score maps, and a decision tree (DT) classifier is employed to segment tumor and sub-tumor parts, producing the segmented sub-tumor regions.

Li et al. (2019) [5] proposed an extension to the U-Net segmentation model by introducing a new cross-layer architecture. This modification includes an "up skip connection" to enhance the model's capability in accurate brain tumor segmentation by learning multi-level features. Additionally, an inception module was integrated to augment the model's representation capacity while managing computational complexity, thereby improving visual information capture. For training, a cascading strategy was employed to segment three distinct subregions: complete tumor, tumor core, and enhancing tumor regions. Experimental results demonstrated progressive optimization of segmentation outcomes with this enhanced model.

Zhao et al. (2018) [6] proposed a deep learning model for brain tumor segmentation, integrating Fully Convolutional Neural Networks (FCNNs) and Conditional Random Fields (CRFs). This model operates on MRI modalities such as T1c, T2, and FLAIR, aiming to classify five distinct classes: healthy tissue, edema, non-enhancing and enhancing cores, and necrosis. The FCNNs consist of two branches receiving inputs of different sizes, which are trained simultaneously. Larger inputs are transformed into feature maps matching the size of smaller inputs. These feature maps are then combined and fed into subsequent networks. The CRF-RNN leverages predictions from FCNNs to refine segmentation labels, optimizing both the appearance and spatial consistency of segmented tumors pixel-by-pixel across image slices.

Chen et al. (2019) [7] introduced a novel deep convolutional symmetric neural network (DCSNN) based on the Baseline Network, akin to Feature Pyramid Network and U-Net architectures. The Baseline Network incorporates lateral connections between blocks across both top-down and bottom-up pathways, facilitating feature fusion. These pathways employ convolutional layers and ResBlocks, which enlarge the receptive field. The top-down pathway uses Deconvolution to refine coarse-grained features into fine-grained ones. In the modified DCSNN, a Left-Right Similarity Mask (LRSM) is integrated selectively into ResBlocks and Deconvolution layers, enhancing lateral connectivity. This network processes four image modalities to synthesize comprehensive asymmetrical features within the LRSM. Comparative evaluation against two Siamese-based methods demonstrates the model's enhanced performance.

Havaei et al. (2016) [8] proposed a two-path cascaded architecture for glioma segmentation. One path focuses on extracting local details while the other emphasizes larger details. This architecture, called TwoPathCNN, concatenates feature maps from both paths and performs class label prediction at the output layer. They also introduced InputCascadeCNN, a cascaded architecture that incorporates local dependencies by using estimates from TwoPathCNN for classification. Alongside these architectures, they introduced the Two-Phase Patch-Wise training procedure, which optimizes training time efficiency.

Havaei et al. 2017 [9] have modified this architecture by introducing a local pathway concatenation which establishes a connection between the first CNN to the second through a hidden layer. Along with this local concatenation, another model has been designed by concatenating the first CNN's output layer with the pre-output layer of the second CNN thereby forming another cascaded architecture with pre-output concatenation.

Chen et al. 2019 [10] proposed a three-dimensional U-Net architecture (S3D-UNet) that utilizes the S3D convolution block in place of a conventional convolution block, replacing the usual 3D convolutions with spatiotemporal-separable 3D convolutions, thereby reducing memory demand as well as computational cost.

Ahmad et al. 2021 [11] have proposed a densely connected 3D U-Net model composed of two types of blocks: dense blocks, which play a vital role in both the encoder-decoder paths of the model and residual-inception blocks, present in the encoder path with the first dense block and in the decoder path's upsampling layer. This model combines the advantages of residual and dense connections with the help of Atrous Spatial Pyramid Pooling (ASPP). By utilizing dense connections and increasing the maximum feature size to 32 in the output layer, the number of features is doubled compared to the U-Net model.

# 3. PROPOSED SYSTEM ARCHITECTURE

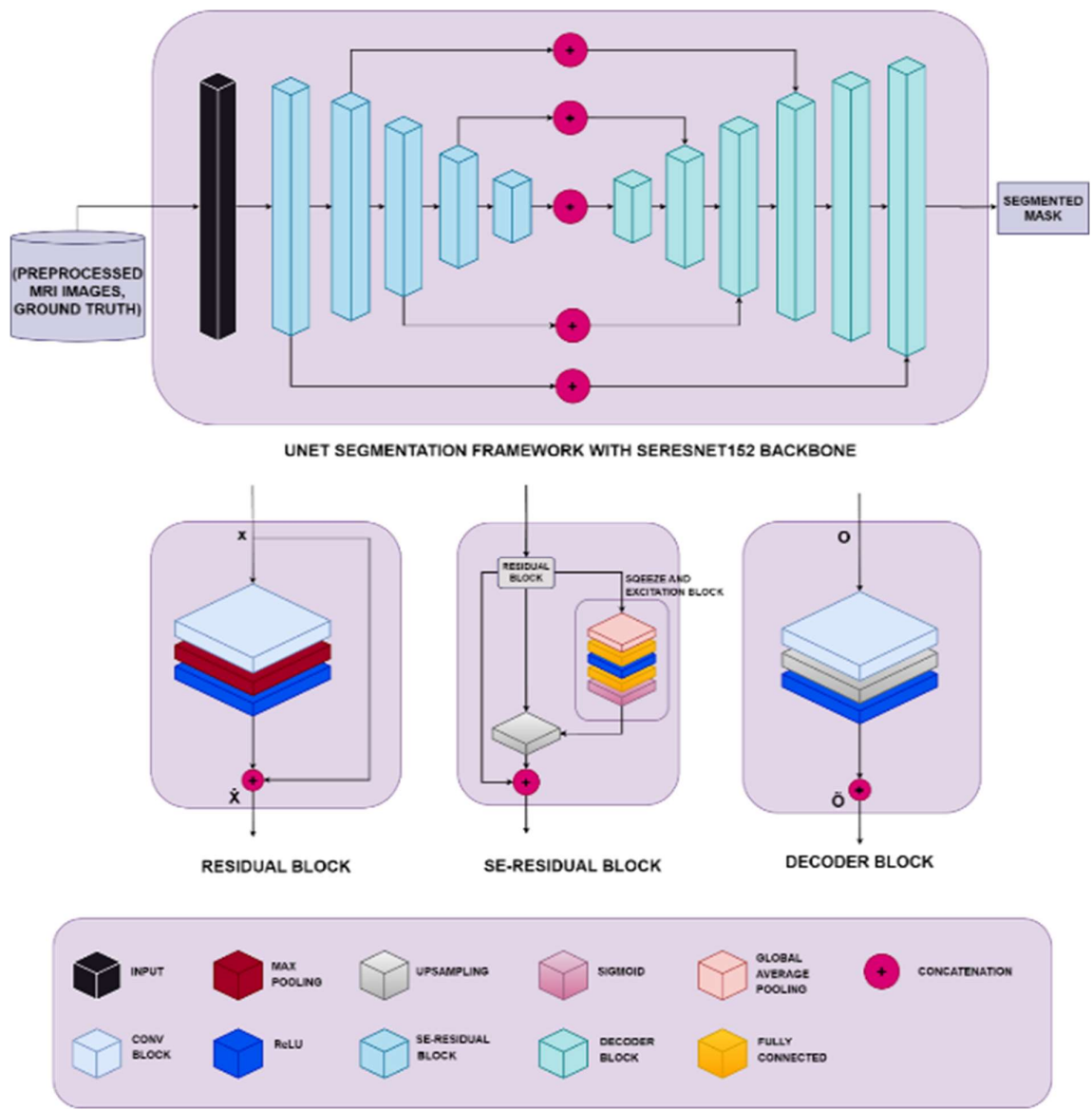

Figure 1. Segmentation System Architecture for BraTS 2020 MRI Brain Tumor Segmentation

## 3.1 DATASET INFORMATION

The MRI image dataset utilized for this experiment is taken from the Brain Tumor Segmentation Challenge 2020 (BraTS 2020) [16-20], which consists of 3D MRI images from multiple institutions focused on segmenting brain tumors, specifically gliomas, from the overall brain image.

This dataset contains a total of 494 different MRI samples, each captured as 3D images in 4 modalities: T1, T1ce, T2, and FLAIR. Out of the 494 sample images in the dataset, 369 images are used during the training phase, while 125 images are utilized for validation. T1ce, T2, and FLAIR images are used to train the proposed segmentation model in this research.

## 3.2 FEATURE EXTRACTION AND SEGMENTATION

### 3.2.1 SeResNet152 CNN Backbone

The SeResNet152 model is a modification of the ResNet-152 architecture, featuring the integration of Squeeze-and-Excitation (SE) blocks. [12] The residual connections in the residual blocks enable the model to effectively utilize numerous layers without encountering the vanishing gradient problem.

A notable augmentation in SeResNet152 involves the inclusion of SE blocks, which dynamically recalibrate feature responses at the channel level by explicitly modelling dependencies among channels. These blocks execute two primary operations: squeeze, which consolidates feature maps across spatial dimensions to generate a channel descriptor, and excitation, which captures inter-channel dependencies and adjusts the feature maps proportionally. This is why we decided to use this model as the backbone of our U-Net segmentation framework.

### 3.2.2 U-Net Framework

The U-Net architecture, an encoder-decoder-based convolutional neural network (CNN), is specialized for semantic image segmentation tasks, particularly in biomedical imaging. It is composed of an encoder, decoder, bottleneck, and skip connections in between the parallel blocks of the encoder and decoder. The encoder in U-Net captures hierarchical features by employing a contracting path, facilitating the extraction of high-level representations. On the other hand, the decoder reconstructs spatial information through an expansive path, generating a segmentation map. Skip connections between the encoder and the corresponding decoder block play a vital role by preserving fine-grained details and aiding gradient flow. The bottleneck in U-Net serves to facilitate the transition between the contracting and expansive paths, capturing critical features in the process.

In the proposed model, the SeResNet152 backbone has been used to function as the encoder. The input preprocessed MRI image would first pass through the initial layers of the backbone, which include convolution layers, a ReLU activation function, and batch normalization.

As the image progresses deeper into the network, it goes through several residual blocks. The SeResNet152 variant includes Squeeze-and-Excitation blocks inside these residual blocks, where the feature maps are adaptively recalibrated. Throughout the model, spatial resolution is reduced (downsampled), which aids in learning increasingly abstract representations.

Since the SeResNet152 model has been used as the encoder backbone, the decoder is also designed to synchronize in parallel with the encoder. The decoder employs upsampling layers to increase the spatial dimensions of the feature maps, counteracting the downsampling effect caused by the encoder's pooling layers. At each upsampling step, the upsampled features are combined with corresponding features from the encoder using skip connections. After merging these features, the decoder applies additional convolutional layers to refine them. These layers reduce the number of feature channels while preserving spatial dimensions, aiming to reconstruct finer details lost during encoding. Finally, the refined features pass through a ReLU activation function to introduce non-linearity.

At the end of the decoder part of the framework, the final convolutional layer is responsible for transforming the segmentation map so that it contains the same number of image channels as the ground truth mask.

## 4. RESULTS AND DISCUSSION

### 4.1 EXPERIMENTAL RESULTS

In our research, we used a U-Net Framework with SeResNet152 as its backbone. This model was trained using the 3D MRI images for 100 epochs and their respective masks from the BraTS 2020 Dataset

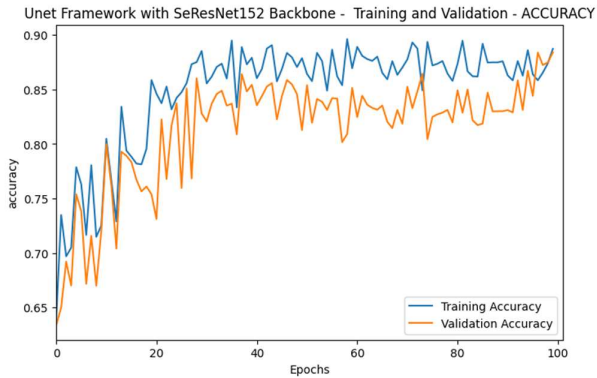

Figure 2. Accuracy Curve for the proposed model

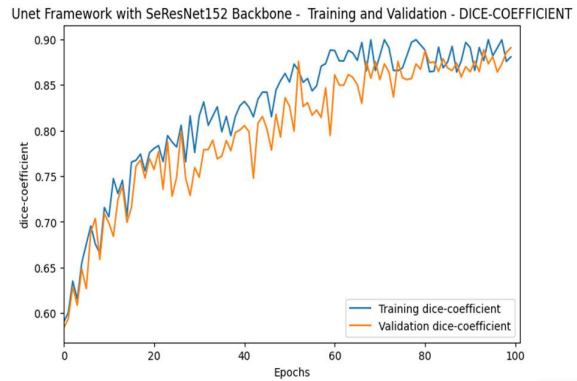

Figure 3. Dice Coefficient Curve for the proposed model

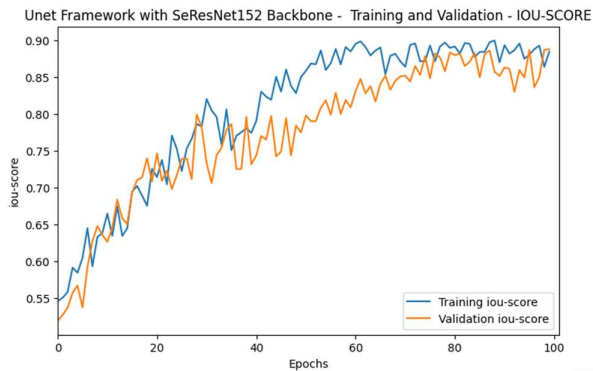

Figure 4. IoU Score Curve for the proposed model

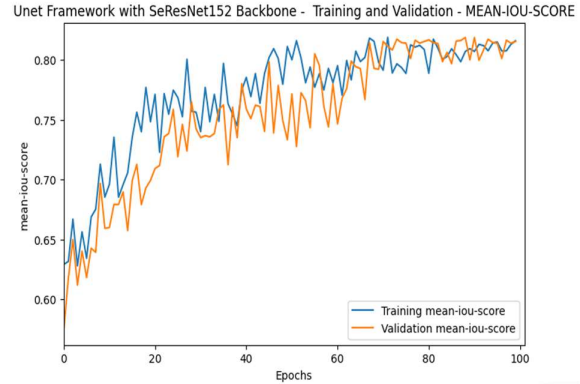

Figure 5. Mean IoU Score curve for the proposed model

The U-Net model demonstrated commendable performance on the BraTS 2020 dataset, particularly in the task of image segmentation. The proposed model's performance was evaluated using various quantitative indicators such as the Dice coefficient, IoU score, and Mean IoU score, which were observed as 0.87, 0.88, and 0.82 respectively. These metrics provide a comprehensive understanding of the model's effectiveness in accurately segmenting the images from our dataset.

## 4.2 COMPARATIVE ANALYSIS

The Dice Coefficient obtained by evaluating the proposed model is put into comparison with other models such as the AMMGS by Liu et al. 2023 [13], Encoder-Decoder based architecture with variational auto-encoder by Myronenko et al. 2018, and Simple Linear Iterative Clustering (SLIC) approach proposed by Iqbal et al. 2022. The results are tabulated in Table 1.

**Table 1. Performance on BraTS 2020 Dataset**

| Model | Dice Coefficient |
|---|---|
| AMMGS [13] | 0.8172 |
| Encoder-Decoder with VAE [14] | 0.8154 |
| SLIC [15] | 0.8593 |
| **Our Proposed Model** | **0.8726** |

Our proposed model surpasses all existing reference models in its performance evaluated in terms of the Dice Coefficient value. It offers a significant advancement in the field, providing a higher level of accuracy, efficiency, and versatility. These results validate the effectiveness of the novel approach and underscore the potential of this model in contributing to a broader body of knowledge.

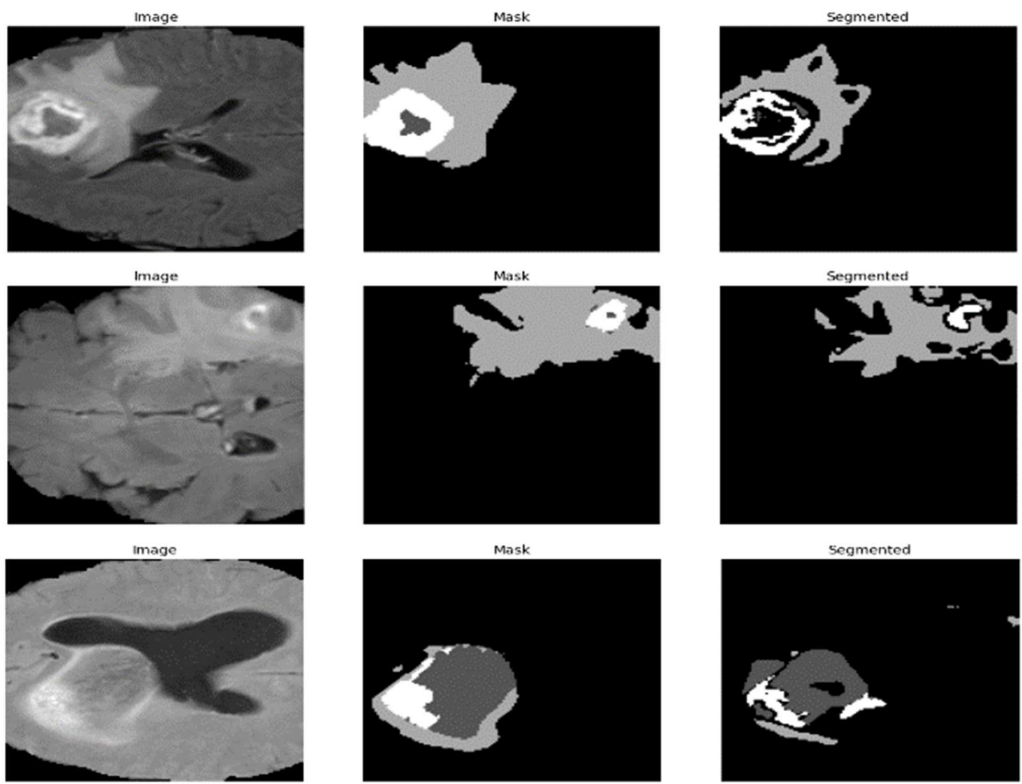

**Figure 6. Segmentation Results on BraTS 2020 Dataset Using the Proposed U-Net Framework with SeResNet152 Backbone**

## 5. CONCLUSION

In conclusion, our U-Net model has demonstrated promising results for segmenting tumors from 3D MRI scans. The model's architecture, characterized by an encoder for downsampling and a decoder for upsampling, enabled efficient feature extraction and precise segmentation. The model's performance was evaluated using various metrics, providing a comprehensive understanding of its effectiveness. Our model achieved an accuracy of 89.12%, a Dice Coefficient of 0.87, an IoU Score of 0.88, and a Mean IoU Score of 0.82. With these metrics, we verify our model's ability to learn robust features, making it an effective solution for segmenting brain tumors from MRI scans.